\documentclass[floatfix,%
 reprint,
 amsmath,amssymb,
 aps,
]{revtex4-2}

\usepackage{dcolumn}
\usepackage{bm}
\usepackage{amsmath}
\usepackage{amssymb}
\usepackage{braket}
\usepackage{subcaption}
\usepackage{graphicx}
\usepackage{caption}
\usepackage{float}
\usepackage{array}
\usepackage{booktabs}
\usepackage{rotating}
\usepackage{tikz}
\usetikzlibrary{shapes, arrows.meta, positioning}

\begin{document}

\title{Beyond Bell Teleportation:\\Machine-Learned Adaptive Protocols}

\author{Krishnajith C Vinod}
 \altaffiliation{Department of Physics, Government Arts and Science College,
Meenchanda, Calicut, Kerala 673018, India}
\author{N C Randeep}%
 \email{randeepvarnam@gmail.com}
\affiliation{Department of Physics, Government Arts and Science College,
Meenchanda, Calicut, Kerala 673018, India \\
 Department of Physics, SARBTM College, Koyilandy, Calicut, Kerala 673018, India}%

\date{\today}

\begin{abstract}
Quantum teleportation have a central role in quantum information science and allows transferring of an unknown quantum state through entanglement and classical communication. Unfortunately, the interaction with external and internal noise severely affects the quality of teleportation and poses limitations on practical applications of quantum communication networks. In this work, instead of conventional Bell teleportation, we introduce  a Machine Learned adaptive protocol for optimizing multiple components of Quantum Teleportation in order to achieve higher fidelity in various noise environments. In order to demonstrate the performance of the proposed scheme, we study three different noise models, including bit-flip, amplitude damping, and depolarizing noise, both in case of single-qubit and two-qubit channels. As a result, we observe substantial improvement in the teleportation fidelity in comparison to the classical Bell-state teleportation protocol in certain noise conditions. Furthermore, the machine-learned protocol reveals a nontrivial strategy for compensation of decoherence and information losses. In addition, obtained results indicate the flexibility and reliability of the proposed framework for implementing various adaptive quantum communications while shedding light on possibilities of discovery of optimal quantum algorithms by means of automated approaches. 

\end{abstract}

\maketitle


\section{\label{sec:level1}INTRODUCTION}
Quantum Communication is one of the major aspects of quantum information science. Quantum communication deals with the communication and processing of information in terms of quantum systems. Unlike conventional or classical communication which is based on electromagnetic signals and their superposition, quantum communication exploits the quantum properties such as entanglement and superposition to facilitate communication~\cite{qc}. In the past decades, many interesting quantum communication protocols have been introduced laying down a groundwork for quantum communication~\cite{qc}.

For example, the protocols such as \textit{remote state preparation} (RSP)~\cite{rsp1, rsp2, rsp3}, in which an unknown quantum state can be prepared remotely via a combination of classical and quantum communications; \textit{superdense coding}~\cite{sdc1, sdc2}, in which two classical bits can be sent with one qubit with the use of entanglement; and \textit{quantum secret sharing}~\cite{qss1, qss2}, in which information is distributed between several parties such that only certain subsets of parties are able to decode the information. These examples reveal the capabilities of the quantum entanglement to be utilized as an information carrier.

All in all, \textit{quantum teleportation}~\cite{qt,qtexp} plays a particularly central role among the above protocols. It provides a scheme for transmission of an unknown quantum state between two places without transferring physical objects.

It is important to note that for the successful practical implementation of quantum teleportation, the effects of external noise become crucial. The effects of the noise make the state of entangled particles worse leading to lower fidelity. Therefore, the key issue here is not just how to implement quantum teleportation, but how to develop the protocols which would still perform perfectly in noisy environment~\cite{noise}. Consequently, the study of adaptation and optimization-based approaches becomes particularly valuable in such context.

In the current work, an optimization algorithm is applied, which incorporates some basic concepts from the machine learning theory~\cite{not,qml1,qml2}. This approach optimizes the parameters of the quantum teleportation protocol aiming at increasing the average fidelity of quantum teleportation subject to certain standard noises.

The outline of the current work goes as follows: Section II provides a brief description of the relevant concepts of machine learning that are used in the present work. Section III provides a brief description of the quantum teleportation protocol in the presence of noise. Section IV provides a detailed description of the mathematical framework of the proposed method and provides a description of the adaptive quantum teleportation protocol and the performance of the learning algorithm. Section V is where we analyze and study the obtained results involving two methodologies 1. Using only the adaptive rotated measurement basis and 2. Fully Adaptive protocol and in Section VI we conclude the paper by summarizing the main results obtained in the present work and discussing possible future directions and potential extensions of the proposed framework.
\section{\label{sec:level1}Review of Machine Learning}
In accordance with the outlined scheme, we start with presenting the fundamental ideas in machine learning that are relevant to our framework. Machine learning is a set of techniques that can optimize complex systems based on data or feedback. In this study, we concentrate on the iterative adjustment of parameters to maximize an objective function. Examples of such methods include the Gradient Ascent method and the Reinforcement Learning method. Gradient based techniques can help achieve this goal systematically by moving along the path of maximum improvement, whereas reinforcement learning can be utilized to make decisions based on trial and error.
\subsection{Gradient Ascent}

Gradient Ascent~\cite{gdal} is an optimisation technique used when the objective is to maximize a function. 
In many mathematical problems, when we want to find the maximum or minimum of a function, we study its gradient. 
The gradient shows the direction in which the function increases most rapidly.

If the goal is to maximize a function, the procedure is called \textit{Gradient Ascent}. If the goal is to minimize a function, it is called \textit{Gradient Descent}~\cite{gdd}.

In the present problem, the quantity we aim to maximize is the teleportation fidelity. The fidelity can be expressed as a function of multiple parameters,

\begin{equation}
F = F(x_1, x_2, \dots, x_{15}),    
\end{equation}

where the parameters are the variables involved in the teleportation protocol, including the entangled quantum channel coefficients, rotation parameters of the measurement basis, and the Post-Processing Quantum Channel parameters.

The goal here is to determine the values of these parameters for which the fidelity becomes maximum.

To do this optimisation, the gradient of the fidelity function with respect to each parameter is calculated,

\begin{equation}
\nabla F =
\left(
\frac{\partial F}{\partial x_1},
\frac{\partial F}{\partial x_2},
\dots,
\frac{\partial F}{\partial x_{15}}
\right).
\end{equation}

Once the gradient is known, the parameters are updated according to the rule

\begin{equation}
x_{\text{new}} = x_{\text{old}} + \eta \nabla F,
\end{equation}

where \( \eta \) is called the \textit{learning rate}. The learning rate determines the size of the step taken in the direction of increasing fidelity.

This update process is performed repeatedly for all parameters simultaneously so that the fidelity increases at every step,

\begin{equation}
F_{\text{new}} > F_{\text{old}}.
\end{equation}

However, applying Gradient Ascent directly to this problem becomes challenging. The teleportation fidelity depends on a complicated sequence of quantum operations, including tensor products, noise channels, measurement operators, and partial trace operations. As a result, calculating analytical gradients with respect to all fifteen parameters becomes computationally cost-heavy.

Another difficulty is that the fidelity function is generally \textit{non–convex}. In such functions, multiple local maxima might exist, and the Gradient Ascent algorithm may find a local maximum instead of the global maximum. Therefore, the results obtained from this method might not be always reliable. For these reasons, we seek an alternative method for the given task.

\subsection{Reinforcement Learning}

Since the optimisation problem involves searching a parameter space of fifteen dimensions, a more reliable machine learning strategy is required. Therefore, instead of using gradient–based optimisation, a reinforcement learning~\cite{rln} style optimisation approach is employed.

The basic idea of this method is to introduce small random variations to the parameters and evaluate the effect of these variations on the teleportation fidelity. If the new set of parameters produces a higher fidelity than the previous one, the algorithm accepts the update. Otherwise, the modification is rejected.

Mathematically, if the fidelity is represented as a function \(F(x)\), where \(x\) represents the set of parameters, the algorithm proposes a perturbed parameter

\begin{equation}
x_{\text{new}} = x_{\text{old}} + \alpha \lambda ,
\end{equation}

where \(\lambda\) represents a random perturbation vector and \(\alpha\) is called the \textit{step size}.

After computing the fidelity using the disturbed parameters, the update rule becomes

\begin{equation}
\text{if } F_{\text{new}} > F_{\text{old}}, \quad x_{\text{old}} \leftarrow x_{\text{new}}.
\end{equation}

This process is done repeatedly until the fidelity approaches its theoretical upper bound.

Conceptually, this procedure is similar to perturbation methods where small variations are introduced to search nearby solutions in the parameter space.

The parameter \(\alpha\), known as the step size, determines the magnitude of the perturbation. The smaller the step size, the slower the progress in the parameter space towards the optimal solution. Thus, convergence will take a long time, making the process computationally expensive. However, a step size that is too large results in random jumps around the parameter space and thus will not be able to converge at all.

To balance exploration and convergence, a step decay mechanism is introduced. At the beginning of the optimisation process, the step size is relatively large, allowing the algorithm to search the parameter space widely. As the algorithm approaches a good solution, the step size gradually decreases, allowing finer adjustments near the best point.

Because the optimization problem involves searching over a fifteen–dimensional parameter space, reinforcement learning based optimisation provides a practical and efficient method for discovering the parameter configuration that maximizes teleportation fidelity.
\section{\label{sec:level1}Review of Noisy Teleportation}
In order to build on the Reinforcement learning framework mentioned above we need to understand how noisy systems function. Quantum teleportation achieves unit fidelity only under perfect ideal conditions. In practice, however, interactions with the environment, imperfect operations, and decoherence introduce noise that reduces the entangled resource shared between the communicating parties. Consequently, the state reconstructed at the receiver varies from the expected input state.
Understanding how noise affects teleportation is therefore essential for assessing the accuracy of the protocol. It also provides a framework for quantifying performance loss and motivates the development of strategies for noise reduction. In what follows, we analyze standard noise models and introduce the required tools to describe their action on quantum states. Subsequently, we study the effect of noise on single-qubit teleportation as an example.

\subsection{Kraus Operators and Standard Noise Models}

The evolution of an open quantum system is conveniently described using the operator-sum (Kraus) representation. Any completely positive trace-preserving (CPTP) map can be written in terms of a set of operators \( \{E_k\} \) satisfying
\begin{equation}
 \sum_k E_k^\dagger E_k = I.   
\end{equation}

For an initial state \( \rho \), the action of the quantum channel is given by
\begin{equation}
 \mathcal{E}(\rho) = \sum_k E_k \rho E_k^\dagger.  
 \label{Eq.quantumevolution} 
\end{equation}
This representation captures the effect of environmental interactions as a probabilistic mixture of different transformations. These Kraus operators are used here to construct noise models which are mentioned in the next section.
\par\vspace{3pt}
\textbf{Standard Noise Models}: Several standard noise models are commonly used in quantum information theory, each described through a set of operators \( E_k \), where \( k \in \{0,1,2,\dots\} \), known as Kraus operators, which capture how a quantum state evolves in the presence of environmental disturbances. Commonly used standardized noise models are
\begin{enumerate}
    \item  \textit{Bit-flip channel:} With probability \( p \), the computational basis states are exchanged \( |0\rangle \leftrightarrow |1\rangle \). Which are represented by Kraus operators,
    \begin{equation}
    E_0 = \sqrt{1 - p} \, I, \quad
    E_1 = \sqrt{p} \, X.
    \label{Eq.btpnoise}
        \end{equation}

    \item \textit{Phase-flip channel:} Introduces a phase error via the Pauli-\( Z \) operator with probability \( p \). Its Kraus operators are
    \begin{equation}
    E_0 = \sqrt{1 - p} \, I, \quad
    E_1 = \sqrt{p} \, Z.
    \label{Eq.pfcnoise}
    \end{equation}

    \item \textit{Depolarizing channel:} With probability \( p \), the qubit is replaced by a maximally mixed state:
     \begin{equation}
    \mathcal{E}_{\text{dep}}(\rho) = (1 - p)\rho + \frac{p}{3}(X\rho X + Y\rho Y + Z\rho Z).
    \label{Eq.dpcnoise}
     \end{equation}

    \item \textit{Amplitude damping channel:} Which describes a model with energy dissipation, such as spontaneous emission. Its Kraus operators are
    \begin{equation}
    E_0 =
    \begin{pmatrix}
    1 & 0 \\
    0 & \sqrt{1 - p}
    \end{pmatrix}, \quad
    E_1 =
    \begin{pmatrix}
    0 & \sqrt{p} \\
    0 & 0
    \end{pmatrix}.
    \label{Eq.ADCnoise}
    \end{equation}
\end{enumerate}

Here, \( p \in [0,1] \) for all four noise models. These Kraus operators, together with the evolution equation  Eq.~(\ref{Eq.quantumevolution}), describe the effect of noise on a single-qubit system. Which can be extended to composite (multi-qubit) systems.  
\par\vspace{3pt}
\textbf{Kraus Operators for Composite Systems:}
To model noise acting on a quantum state, the operator-sum representation is applied. For composite systems, when noise affects only one subsystem, the corresponding Kraus operators are tensored with the identity acting on the unaffected subsystem:
\begin{equation}
\mathcal{E}(\rho) = \sum_k (I \otimes E_k)\,\rho\,(I \otimes E_k^\dagger).
\label{Eq.sngentnoise}
\end{equation}

More generally, if noise acts on both subsystems of an entangled state, the evolution is described by independent Kraus operators applied to each qubit:
\begin{equation}
\mathcal{E}(\rho) = \sum_{k,\ell} (E_k \otimes E_\ell)\,\rho\,(E_k^\dagger \otimes E_\ell^\dagger),
\label{Eq.dobentnoise}
\end{equation}
where \(E_k\) and \(E_\ell\) correspond to the noise processes acting on the respective subsystems.

The resulting density matrix contains the full effect of the interactions with the environment and can be used for further analysis, e.g., for evaluating the result of a measurement or the fidelity of the process. 

The standard noise models discussed above can be applied to a quantum system subjected to noise. Here, we are interested in single-qubit teleportation in the presence of noise.

\subsection{Noise in Single Qubit Teleportation}
For single-qubit teleportation, the state to be teleported is the input state,
\begin{equation}
\ket{\psi} = a_{0} \ket{0} + a_{1} \ket{1}, \quad |a_{0}|^2 + |a_{1}|^2 = 1.
\end{equation}

We introduce various noise models using the Kraus operators discussed in Eqs. (\ref{Eq.btpnoise}), (\ref{Eq.pfcnoise}), (\ref{Eq.dpcnoise}), and (\ref{Eq.ADCnoise}). The resulting noisy state is then given by

\begin{equation}
\rho_{\text{noisy}} = \sum_{i=0}^1 E_i \ket{\psi}\bra{\psi} E_i^\dagger.
\end{equation}
then the remaining process remain the same as that of standard Bell Teleportation. For a noise model acting on the entangled channel, evolve the respective density matrix $\rho_{ent}$ according to Eq. (\ref{Eq.sngentnoise}) or Eq. (\ref{Eq.dobentnoise}). After obtaining the final state, the next step is to calculate the fidelity of teleportation.
\par\vspace{3pt}
\textbf{Fidelity Calculation:}
The fidelity can be defined as the measure of success achieved when a quantum state is reproduced or transmitted. Fidelity has a numerical value that varies between 0 and 1. 

When fidelity equals 1, it signifies perfect transmission. When fidelity equals 0, it shows there was a total loss of information during the transmission process. 

Let the input state be
\[
\ket{\psi(\alpha,\beta)} = \cos\alpha \ket{0} + e^{i\beta}\sin\alpha \ket{1},
\]
with corresponding density matrix
\[
\rho_{\text{in}} = \ket{\psi}\bra{\psi}.
\]

Then, the total input state during teleportation is given by $\rho_{total}=\rho_{in} \otimes \rho_{ent}$. After the teleportation protocol, conditioned on measurement outcome \(i\), the output state is \(\rho_i\), obtained with probability
\[
p_i = \mathrm{Tr}\big(M_i \rho_{\text{total}}\big),
\]
where \(M_i\) are the measurement operators.

The fidelity for outcome \(i\) is
\begin{equation}
F_i(\alpha,\beta) = \bra{\psi} \rho_i \ket{\psi}.
\end{equation}

The total fidelity for a given input state is the probability-weighted sum
\begin{equation}
F(\alpha,\beta) = \sum_{i=1}^{4} p_i \, F_i(\alpha,\beta).
\end{equation}

To evaluate the performance of the protocol, we average over all input states on the Bloch sphere:
\begin{equation}
F_{\text{avg}} = \frac{1}{4\pi} \int_{0}^{\pi} \int_{0}^{2\pi}
F(\alpha,\beta) \, \sin\alpha \, d\beta \, d\alpha.
\label{Eq.AvgFdltBlochsphere}
\end{equation}
This is the average fidelity of teleportation for all possible input states, which determines how our protocol is affected by the noise parameter p.
\section{Application of Machine Learning Techniques in Noisy Quantum Teleportation}

In this section, we replace the original Bell teleportation protocol with an adaptive scheme governed by a machine learning technique. In the adaptive scheme, the basic teleportation procedure is adapted by adding learnable parameters in the entanglement channel, measurement basis, and correction procedure. The aim is to tune these parameters to achieve maximum fidelity. Before proceeding, we first define the adaptive parameters used in our scheme, and then discuss the role of machine learning techniques in tuning these parameters.
\subsection{Adaptive Parameters}
Teleportation can be successfully achieved through a series of steps and is influenced by several factors. These factors are related through specific parameters and influence the fidelity of teleportation in the presence of noise. We made these parameters tunable to improve fidelity and treated the following parameters as adaptable.
\subsubsection{Adaptive Entangled Channel}

Instead of a fixed Bell state, the shared entangled resource is taken as a general two-qubit state:
\begin{equation}
\ket{\Phi} = a\ket{00} + b\ket{01} + c\ket{10} + d\ket{11},
\end{equation}
where $a,b,c,d \in \mathbb{C}$ satisfy the normalization condition
\[
|a|^2 + |b|^2 + |c|^2 + |d|^2 = 1.
\]
These four parameters can be considered adaptive parameters of the entangled channel. The corresponding density matrix is then given by
\[
\rho_{\text{ent}} = \ket{\Phi}\bra{\Phi}.
\]

This state is subjected to noise acting on Alice’s qubit:
\[
\rho_{\text{ent}} \rightarrow \sum_k (E_k \otimes I)\, \rho_{\text{ent}} \,(E_k^\dagger \otimes I).
\]
This adaptively noise-introduced state is used for the remaining steps of the teleportation protocol.

\subsubsection{Adaptive Measurement Basis}

The measurement is performed on the two-qubit subsystem belonging to Alice, consisting of the unknown input qubit and her part of the entangled pair. This joint measurement is carried out in a rotated Bell basis.

We first construct a reference state by applying a general single-qubit unitary transformation $U(\phi,\theta,\lambda)$ on one qubit of the Bell state
\begin{equation}
\ket{\Phi^+} = \frac{1}{\sqrt{2}}(\ket{00} + \ket{11}),
\end{equation}
giving
\begin{equation}
\ket{B_0} = (U \otimes I)\ket{\Phi^+}.
\end{equation}

The parameters $\phi, \theta$ and $\lambda$ are the adaptive parameters introduced in the measurement basis. The remaining three basis states are then generated by applying Pauli operators $P_i \in \{I, Z, X, ZX\}$ on the same qubit:
\begin{equation}
\ket{B_i} = (P_i \otimes I)\ket{B_0}.
\end{equation}

The corresponding measurement operators act on Alice’s two-qubit subsystem and are given by
\[
M_i = \ket{B_i}\bra{B_i} \otimes I,
\]
where the identity operator acts on the receiver’s qubit, indicating that the measurement does not affect that part of the system. 

\subsubsection{Adaptive Correction Operations}

At the final stage of teleportation, the receiver performs a unitary operation belonging to the Pauli-operator set based on the sender's classical information. To make the protocol adaptive, instead of fixed Pauli corrections, Bob applies general unitary operations:
\begin{equation}
\rho_i \rightarrow U_i \rho_i U_i^\dagger,
\end{equation}
where each $U_i$ is parameterized as a general unitary:
\begin{equation}
U_i = U(\phi_i,\theta_i,\lambda_i).
\end{equation}
The parameters $\phi_{i}, \theta_{i}$ and $\lambda_{i}$ are the adaptive parameters introduced in the correction operators.
\subsubsection{Post-Processing Correction}

To further reduce noise, a post-processing channel is applied to the corrected state:
\begin{equation}
\rho_{\text{corr}} \rightarrow \sum_j J_j \rho_{\text{corr}} J_j^\dagger,
\end{equation}
where the Kraus operators $\{J_j\}$ satisfy
\begin{equation}
\sum_j J_j^\dagger J_j = I.
\end{equation}
We made these Kraus operators $J_j$ adaptive to achieve maximum fidelity for a particular system.

Once we have defined all the adaptive parameters, it is time to define the objective function of this protocol. That is machine learning turned fidelity.

For a given input state $\ket{\psi(\alpha,\beta)}$, the fidelity is
\begin{equation}
F(\alpha,\beta) = \sum_i p_i \, \bra{\psi} \rho_i \ket{\psi}.
\end{equation}

The optimization objective is to maximize the average fidelity:
\begin{equation}
F_{\text{avg}} = \frac{1}{4\pi} \int_0^\pi \int_0^{2\pi}
F(\alpha,\beta)\, \sin\alpha \, d\beta \, d\alpha.
\end{equation}
The structure and Operation of the Unitary $U(\phi,\theta,\lambda)$, The correction operators and Post-Processing Kraus Operators are given in \textbf{Appendix~\ref{AppendixA}} and \textbf{Appendix~\ref{AppendixB}}. 
\par\vspace{3pt}
The formulation outlined above suggests that we have entered a multi-dimensional optimization problem landscape. In particular, our optimization process is subject to various factors including those of the entangled states' parameters, measurement basis rotation angles, correction unitaries, as well as Post-Processing channels. As the problem is noisy and requires multiple interactions of all these factors, an analytical solution is impossible. So, we introduce a data-driven optimization method. Before presenting the results, we discuss the role of machine learning techniques in determining the adaptive parameters for quantum teleportation.
\subsection{Role of Machine Learning}
As mentioned above our aim is to effectively search through the high-dimensional parameter space, The Reinforcement learning module is responsible for optimization within the parameter space.
\begin{equation}
\{a,b,c,d\}, \quad \{\phi,\theta,\lambda\}, \quad \{J_0,J_1\}.
\end{equation}

In order to get the best out of $F_{\text{avg}}$, these parameters are updated over and over again. A new set of parameters is generated at each step, and the teleportation protocol is tested using the noise model used. The average fidelity is checked over the entire Bloch sphere. Based on the improvement in the fidelity achieved, the new set of parameters is accepted or rejected. The algorithm improves the parameter set step by step until the best is achieved.

Now, we first examine the role of certain machine learning techniques, along with their features and limitations, in determining the adaptive parameters for quantum teleportation.
\par\vspace{3pt}
\textbf{Role of Gradient-Based Formulation and Its Limitations:}
The optimization problem can be theoretically arranged under a gradient-based framework. Let $\Theta$ be the collection of all parameters of the protocol that need to be optimized. The objective is to maximize the average fidelity $F_{\text{avg}}(\Theta)$ on the Bloch sphere.

A standard approach would be to employ gradient ascent:
\begin{equation}
\Theta_{t+1} = \Theta_t + \eta \, \nabla_{\Theta} F_{\text{avg}}(\Theta_t),
\end{equation}
where $\eta$ is the learning rate. However, in the present setting, such an approach becomes computationally prohibitive.

The parameter space $\Theta$ is high-dimensional and consists of complex-valued variables subject to normalization and completely positive trace-preserving (CPTP) constraints. Furthermore, the fidelity function is not available in closed analytical form, but is instead evaluated numerically through a discretized integration over the Bloch sphere as discusssed in Eq.~(\ref{Eq.AvgFdltBlochsphere}).

As a result, the computation of the analytical gradients $\nabla_{\Theta} F_{\text{avg}}$ is heavy. Moreover, the computation of the numerical gradients will involve repeated computation of the teleportation protocol for every dimension in the parameters $\Theta$, which is not efficient for a high-dimensional parameter space. Furthermore, the optimization problem is non-convex. So, we introduce another machine learning technique to reduce computational complexity.
\par\vspace{3pt}
\textbf{Role of Reinforcement Learning Approach:}
To address the aforementioned challenges, we use reinforcement learning inspired stochastic optimization strategy. Instead of explicitly computing gradients, the algorithm searches the parameter space through stochastic perturbations of $\Theta$. At each iteration, a candidate parameter set $\Theta'$ is generated and the corresponding fidelity $F_{\text{avg}}(\Theta')$ is evaluated. We now examine the elements and features of the reinforcement learning approach in this adaptive teleportation protocol.

\subsubsection{State and Action Space}
The action space of this model is the set of all trainable parameters, denoted by
\begin{equation}
\Theta = \{a,b,c,d,\ \phi,\theta,\lambda,\ \phi_i,\theta_i,\lambda_i,\ J_j\}.
\end{equation}

At any iteration $t$, the current configuration of the protocol is represented as a state:
\begin{equation}
s_t \equiv \Theta_t.
\end{equation}

An action corresponds to a stochastic perturbation of the parameters:
\begin{equation}
a_t: \Theta_t \rightarrow \Theta_t' = \Theta_t + \epsilon_t,
\end{equation}
where $\epsilon_t$ is a random update sampled from a distribution (typically Gaussian).

\subsubsection{Environment and Transition}

Given a parameter set $\Theta_t$, the environment simulates the full teleportation protocol under noise and produces an output fidelity:
\begin{equation}
s_t \xrightarrow{a_t} s_{t+1} = \Theta_t'.
\end{equation}

The transition is deterministic with respect to the chosen parameters, but stochastic due to the sampling of input states over the Bloch sphere.

\subsubsection{Reward Function}

The reward is defined as the average fidelity of teleportation:
\begin{equation}
R(\Theta) = F_{\text{avg}}(\Theta).
\end{equation}

For numerical implementation, this is approximated using discrete sampling:
\begin{equation}
F_{\text{avg}} \approx \frac{\sum_{\alpha,\beta} \sin\alpha \, F(\alpha,\beta)}{\sum_{\alpha,\beta} \sin\alpha}.
\end{equation}

\subsubsection{Policy and Update Rule}

Instead of an explicit parameterized policy, we employ a stochastic search strategy. A new parameter set $\Theta'$ is accepted if it improves the reward:
\begin{equation}
\Theta_{t+1} =
\begin{cases}
\Theta_t', & \text{if } R(\Theta_t') > R(\Theta_t), \\
\Theta_t, & \text{otherwise}.
\end{cases}
\end{equation}

Additionally, a small probability of accepting suboptimal moves is introduced to encourage exploration:
\begin{equation}
\Theta_{t+1} =
\begin{cases}
\Theta_t', & \text{if } R(\Theta_t') > R(\Theta_t) \\
\Theta_t', & \text{with probability } \eta \\
\Theta_t, & \text{otherwise}.
\end{cases}
\end{equation}

\subsubsection{Learning Dynamics}

The update step size is gradually reduced:
\begin{equation}
\epsilon_t \sim \mathcal{N}(0, \sigma_t^2), \quad \sigma_t \rightarrow 0 \text{ as } t \rightarrow \infty.
\end{equation}

This ensures convergence toward an optimal parameter configuration.

Now, we can examine the workflow of this protocol, that is, how it proceeds and achieves improved teleportation fidelity by introducing adaptive parameters.
\par
\begin{figure}[H]
    \centering
    \includegraphics[width=2\linewidth]{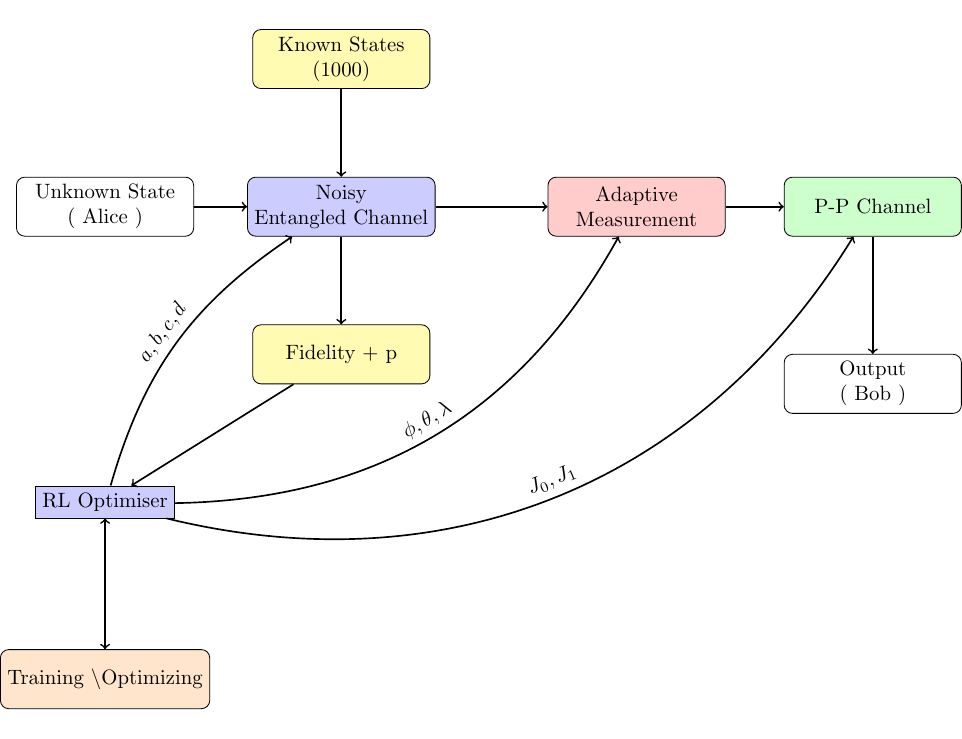}
    \caption{\centering Protocol Workflow of Fully Adaptive Protocol}
    \label{figure1 workflow}
\end{figure}

\vspace{3pt}
\textbf{Protocol Workflow:}
The overall workflow of the proposed protocol is illustrated schematically in FIG.~\ref{figure1 workflow}.The aim is to efficiently teleport an unknown quantum state prepared by Alice through an entangled quantum channel subjected to environmental noise. First during the training stage, a dataset of known quantum states is supplied to the protocol, and the teleportation fidelity obtained for each state serves as the reward signal for the reinforcement learning optimizer. The training dataset consists of 1000 randomly generated single qubit states of the form
\begin{equation}
\ket{\psi} = \cos\alpha \ket{0} + e^{i\beta}\sin\alpha \ket{1},
\end{equation}
where the parameters $\alpha$ and $\beta$ are sampled over 500 distinct values each to ensure broad coverage of the Bloch sphere. For every selected value of $\alpha$, the phase parameter $\beta$ is varied systematically, and the corresponding single-shot teleportation fidelity is evaluated for each configuration. These fidelities are subsequently averaged over all values of $\beta$ to obtain the mean fidelity associated with a given $\alpha$. The procedure is independently repeated for all four orthonormal Bell-state measurement outcomes, following which an overall average fidelity is computed. Furthermore, the noise parameter $p$ is sampled over 50 uniformly distributed values within the interval $[0,1]$, and for each value of $p$, the optimization algorithm is executed for 3000 iterations. During the optimization process, the learning step size is updated using a multiplicative decay schedule with a decay factor of $0.999$ applied at every iteration. Based on the fidelity feedback obtained throughout training, the optimizer iteratively updates the channel coefficients $(a,b,c,d)$, the adaptive measurement parameters $(\phi,\theta,\lambda)$, and the post-processing operators $(J_0,J_1)$ in order to maximize the average teleportation fidelity under the considered noise model.

Thus, we have defined all the tools and techniques required for this protocol, and we now proceed to discuss the results.

\section{Result}

Having that we already identified the learning algorithm as an instrument of constant exploration and optimization of the parameters’ domain, we proceed to analyze its efficiency under realistic conditions. In particular, we consider the averaged fidelity of teleportation under various types of noise for the purpose of testing our adaptive approach. Its efficiency is tested progressively, first under the measurement adaptation and then the full channel optimization.

\subsubsection{\textbf{Only Rotated Basis Scheme}}

In the first stage, only the measurement basis and corresponding correction operations are optimized through the parametrized unitary \( U(\phi,\theta,\lambda) \), while the quantum channel remains unchanged. This allows us to isolate the contribution of measurement adaptation in removing noise effects. Details of the hyper-parameters and datasets used during training are provided in \textbf{Appendix}~\ref{AppendixA}.

The fidelity of this scheme is compared against the standard Bell-state teleportation protocol under three different noise models:
\begin{itemize}
    \item Bit-flip noise
    \item Amplitude damping noise
    \item Depolarizing noise
\end{itemize}
For each noise model, two configurations are considered:
\begin{enumerate}
\renewcommand{\labelenumi}{(\alph{enumi})}
    \item Noise acting only on Alice’s qubit
    \item Noise acting on both qubits of the entangled pair
\end{enumerate}

\begin{figure}[H]
    \caption{Noise on Alice's Bit of Channel}
    \centering
    
    \begin{subfigure}{0.45\textwidth}
        \centering
        \includegraphics[width=\linewidth]{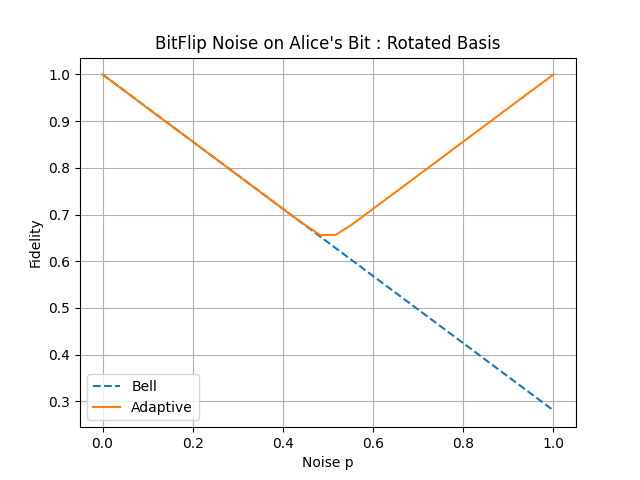}
        \caption{Bitflip Noise on Alice's Bit - Rotated Basis}
        \label{Fig_OnlyRnoisebitflip}
    \end{subfigure}
    \hfill
    \begin{subfigure}{0.45\textwidth}
        \centering
        \includegraphics[width=\linewidth]{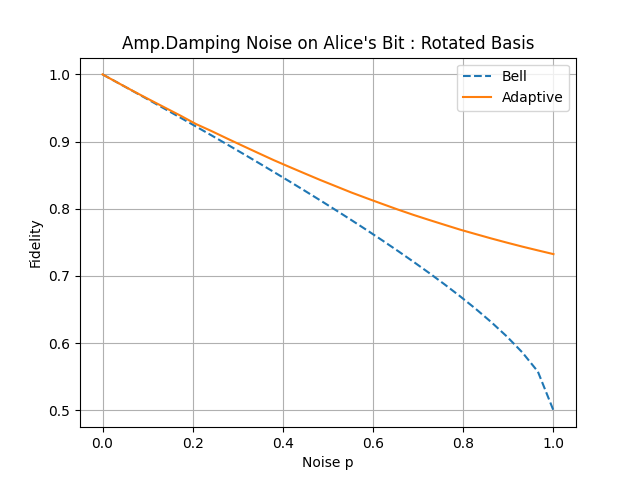}
        \caption{Amplitude Damping Noise on Alice's Bit - Rotated Basis}
        \label{Fig_OnlyRnoiseAD}
    \end{subfigure}
    
    \vspace{0.5cm}
    
    \begin{subfigure}{0.45\textwidth}
        \centering
        \includegraphics[width=\linewidth]{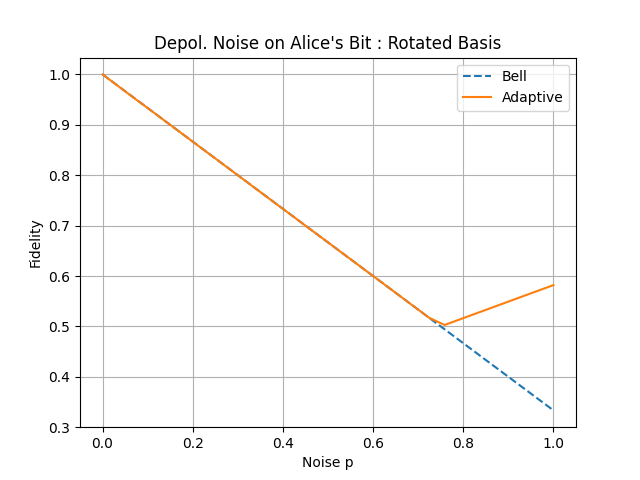}
        \caption{Depol. Noise on Alice's Bit - Rotated Basis}
        \label{Fig_OnlyRnoiseDP}
    \end{subfigure}

\end{figure}

The fidelity variation under noise acting on Alice's parts of the entangled channel for the rotated scheme and standard Bell teleportation is shown in FIG.~ (\ref{Fig_OnlyRnoisebitflip}), (\ref{Fig_OnlyRnoiseAD}) and (\ref{Fig_OnlyRnoiseDP}). In all three cases, the rotated scheme shows a significant improvement in fidelity over standard Bell teleportation.

\begin{figure}[h!]
    \caption{Noise on entire Channel}
    \centering
    
    \begin{subfigure}{0.45\textwidth}
        \centering
        \includegraphics[width=\linewidth]{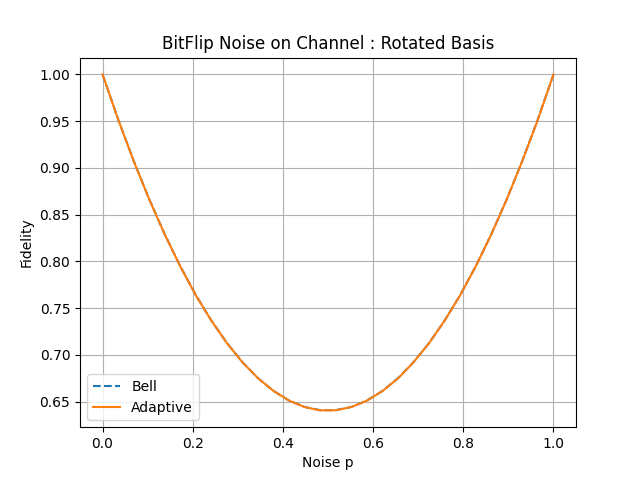}
        \caption{Bitflip Noise on Entangled Channel - Rotated Basis}
        \label{Fig_FullRnoisebitflip}
    \end{subfigure}
    \hfill
    \begin{subfigure}{0.45\textwidth}
        \centering
        \includegraphics[width=\linewidth]{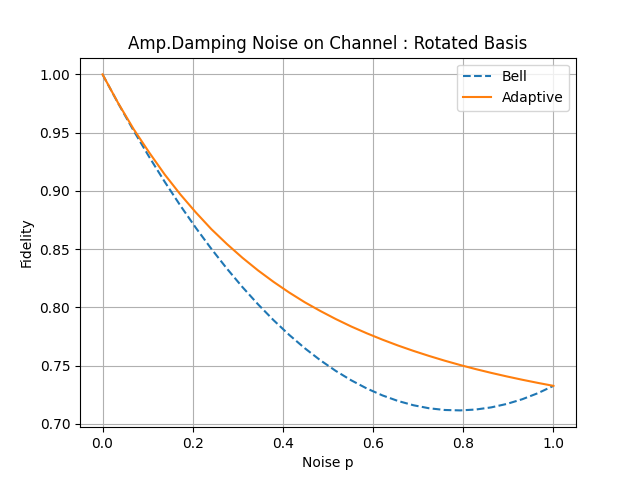}
        \caption{Amplitude Damping Noise on Entangled Channel - Rotated Basis}
        \label{Fig_FullRnoiseAD}
    \end{subfigure}
    
    \vspace{0.5cm}
    
    \begin{subfigure}{0.45\textwidth}
        \centering
        \includegraphics[width=\linewidth]{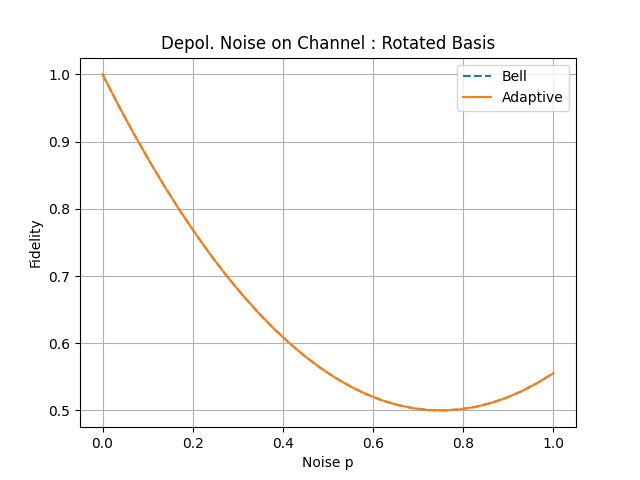}
        \caption{Depol. Noise on Entangled Channel - Rotated Basis}
        \label{Fig_FullRnoiseDP}
    \end{subfigure}

\end{figure}

Similarly, the fidelity variation under noise acting on both qubits of the entangled pair for the rotated scheme and standard Bell teleportation is shown in FIG.~(\ref{Fig_FullRnoisebitflip}), (\ref{Fig_FullRnoiseAD}) and (\ref{Fig_FullRnoiseDP}). Out of the three cases, the amplitude damping cases exhibit improved fidelity in the rotated basis scheme, whereas in the other two cases both methods show the same fidelity variation. For further improvement in fidelity, we make more parameters adaptable.

\subsubsection{Fully Adaptive Protocol}
In the second stage, the protocol is extended to include adaptive quantum channels and post-processing operations in addition to the rotated measurement basis. Specifically, the post-processing Quantum Channel Kraus operators are optimized along with the measurement parameters.

This results in a fully adaptive teleportation scheme, where all components of the protocol are tuned to counteract the effects of noise.

\onecolumngrid
\twocolumngrid

The adaptive parameters appear in the entangled channel,
\[
\ket{\psi} = a\ket{00} + b\ket{01} + c\ket{10} + d\ket{11},
\]
the measurement basis through the unitary $U(\phi,\theta,\lambda)$, and the Post-Processing channel described by the Kraus operator
\[
J_0 = 
\begin{pmatrix}
J0_{00}^{(Re)} + i J0_{00}^{(Im)} & J0_{01}^{(Re)} + i J0_{01}^{(Im)} \\
J0_{10}^{(Re)} + i J0_{10}^{(Im)} & J0_{11}^{(Re)} + i J0_{11}^{(Im)}
\end{pmatrix}.
\]
\[
J_1 = 
\begin{pmatrix}
J1_{00}^{(Re)} + i J1_{00}^{(Im)} & J1_{01}^{(Re)} + i J1_{01}^{(Im)} \\
J1_{10}^{(Re)} + i J1_{10}^{(Im)} & J1_{11}^{(Re)} + i J1_{11}^{(Im)}
\end{pmatrix}.
\]

The optimized parameters are fitted to cubic polynomials (see Table \ref{tab:bitflip_fulladaptive}, \ref{tab:ad_fulladaptive} and \ref{tab:depol_fulladaptive}  in \textbf{Appendix:~\ref{AppendixB}}).

\vspace{3pt}
\par
The underlying motivation for this approach is inspired by earlier works demonstrating that, counter-intuitively, the controlled application of noise or noise-like operations can enhance teleportation fidelity under certain conditions~\cite{fighting_noise,qutrit_noise,qdn}. These studies suggest that appropriately engineered quantum operations can partially compensate for environmental decoherence.

\begin{figure}[H]
    \caption{Noise on Alice's Bit of Channel}
    \centering
    
    \begin{subfigure}{0.45\textwidth}
        \centering
        \includegraphics[width=\linewidth]{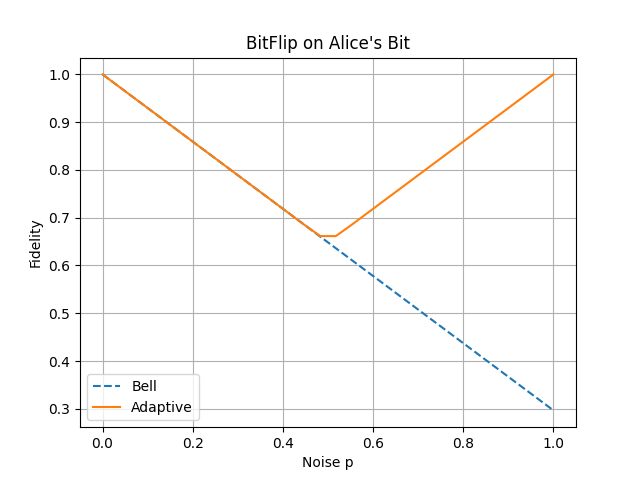}
        \caption{Bitflip Noise on Alice's Bit - Fully Adaptive Protocol}
        \label{Fig_FAPAnoisebitflip}
    \end{subfigure}
    \hfill
    \begin{subfigure}{0.45\textwidth}
        \centering
        \includegraphics[width=\linewidth]{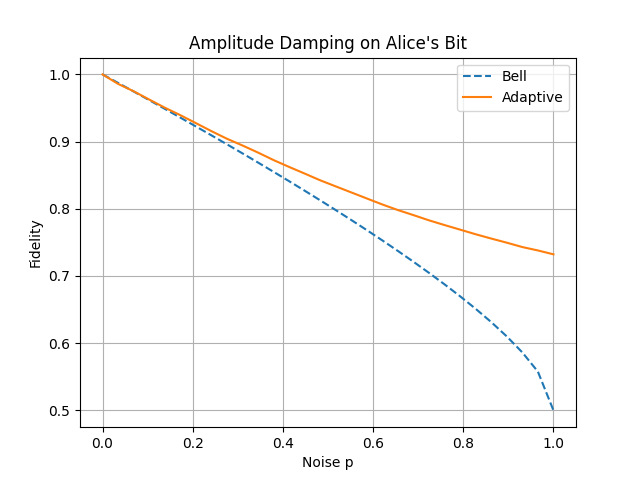}
        \caption{Amplitude Damping Noise on Alice's Bit - Fully Adaptive Protocol}
        \label{Fig_FAPAnoiseAD}
    \end{subfigure}
    
    \vspace{0.5cm}
    
    \begin{subfigure}{0.45\textwidth}
        \centering
        \includegraphics[width=\linewidth]{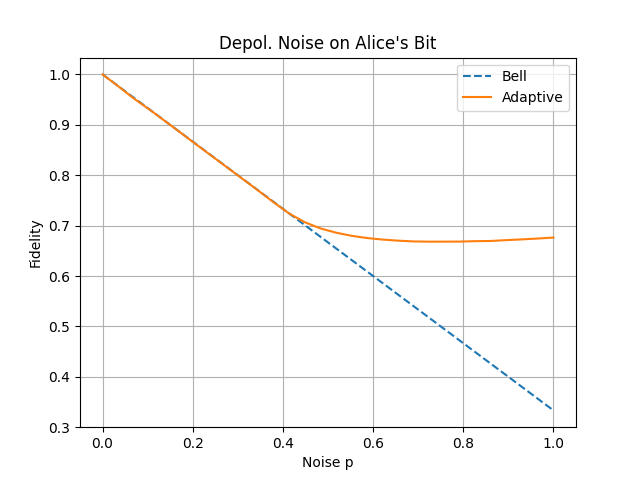}
        \caption{Depol. Noise on Alice's Bit - Fully Adaptive Protocol}
        \label{Fig_FAPAnoiseDP}
    \end{subfigure}

\end{figure}

\begin{figure}[H]
    \caption{Noise on entire Channel}
    \centering
    
    \begin{subfigure}{0.45\textwidth}
        \centering
        \includegraphics[width=\linewidth]{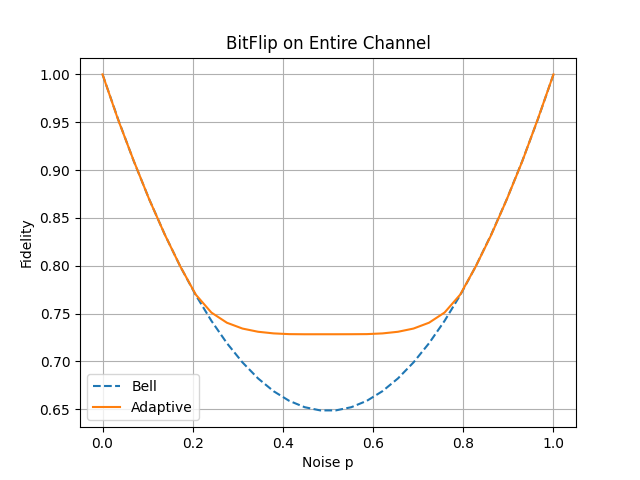}
        \caption{Bitflip Noise on Entangled Channel - Fully Adaptive Protocol}
        \label{Fig_FAPnoisebitflip}
    \end{subfigure}
    \hfill
    \begin{subfigure}{0.45\textwidth}
        \centering
        \includegraphics[width=\linewidth]{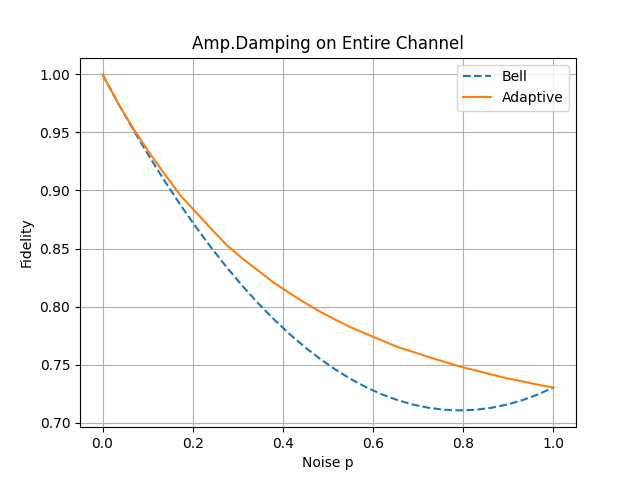}
        \caption{Amplitude Damping Noise on Entangled Channel - Fully Adaptive Protocol}
        \label{Fig_FAPnoiseAD}
    \end{subfigure}
    
    \vspace{0.5cm}
    
    \begin{subfigure}{0.45\textwidth}
        \centering
        \includegraphics[width=\linewidth]{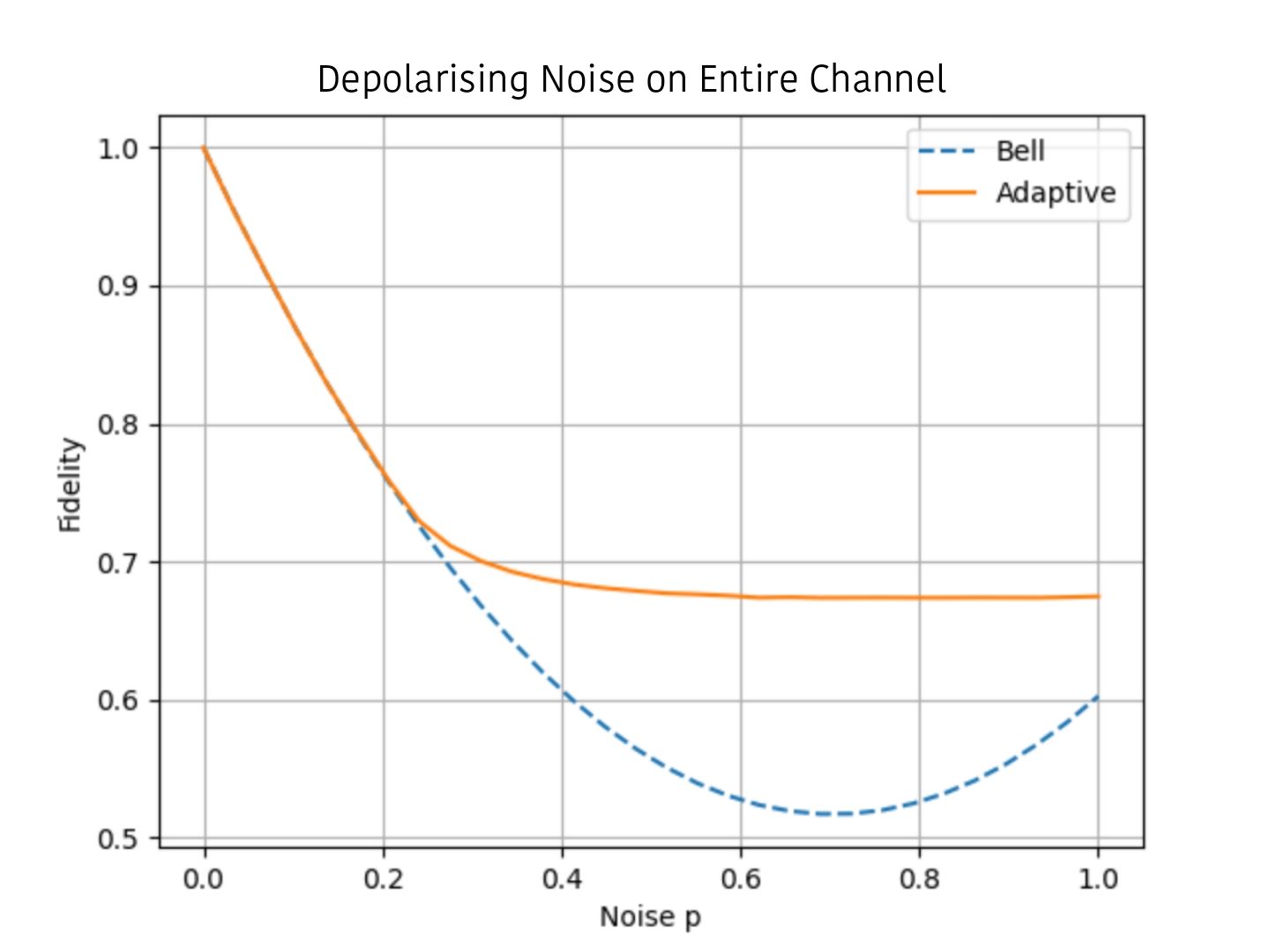}
        \caption{Depol. Noise on Entangled Channel - Fully Adaptive Protocol}
        \label{Fig_FAPnoiseDP}
    \end{subfigure}

\end{figure}

Thus this fully adaptive protocol is then proposed to address the limitation. By optimizing the entangled channel itself, as well as a Post-Processing quantum channel, a significant improvement in fidelity can be achieved see FIG.~ (\ref{Fig_FAPAnoisebitflip}), (\ref{Fig_FAPAnoiseAD}) and (\ref{Fig_FAPAnoiseDP}),  (\ref{Fig_FAPnoisebitflip}), (\ref{Fig_FAPnoiseAD}) and (\ref{Fig_FAPnoiseDP}) This improvement can be seen when there is amplitude damping noise on Alice’s side of the entangled channel as well as moderately high improvement when there is noise on the entire entangled channel for bit flip and depolarizing noise.

\par
These results demonstrate the cumulative advantage of incorporating both measurement adaptation,channel optimization and Post-Processing Channel, leading to significant improvements in teleportation fidelity across a wide range of noise strengths. Thus, we have successfully developed an efficient framework for engineering a quantum teleportation protocol capable of adaptively responding to external noise, thereby maximizing the extractable teleportation fidelity through a Machine Learning based optimization approach.

\vspace{3pt}



\section{Conclusion}
As stated above from our results, it is clear that adaptive protocols developed in this manner can substantially enhance the standard teleportation protocol based on a static Bell measurement and a maximally entangled channel. In particular, our optimized protocol achieve high teleportation fidelity even in situations where the standard protocol exhibits very fast degradation under noisy conditions. This indicates that not only do noisy conditions degrade performance, but also a static protocol may not be an optimal solution.

This is further emphasized by the fact that by making all of the major components of the teleportation protocol adaptive, the protocol can compensate for different types of noise. This underlines the need to view teleportation as an optimizable process. 

It is also interesting to note that by using a learning-based optimization method, a large parameter space can be systematically explored, which may not be easily achieved by using a purely analytical method. This indicates that adaptive protocols could be very important in quantum information science, particularly in quantum communication protocols where noisy conditions play a major role in limiting performance.
\par
We therefore expect that similar techniques could be extended to other tasks in quantum information science, such as quantum error mitigation, channel discrimination, and quantum network optimisation, where noise and imperfect operation are fundamental limiting factors.

\appendix
\section{Adaptive Measurement Basis and Correction Operations \label{AppendixA}}

In the standard teleportation protocol, the sender performs a measurement in the Bell basis, and the receiver applies fixed Pauli corrections. In order to allow the protocol to adapt to noise, we generalize both the measurement basis and the corresponding correction operations.

We introduce a parametrized single-qubit unitary transformation \( U(\phi,\theta,\lambda) \in SU(2) \), defined as
\[
U(\phi,\theta,\lambda) = R_z(\phi)\, R_y(\theta)\, R_z(\lambda),
\]
where the rotation operators are given by
\[
R_z(\alpha) =
\begin{bmatrix}
e^{-i\alpha/2} & 0 \\
0 & e^{i\alpha/2}
\end{bmatrix}, \quad
R_y(\theta) =
\begin{bmatrix}
\cos(\theta/2) & -\sin(\theta/2) \\
\sin(\theta/2) & \cos(\theta/2)
\end{bmatrix}.
\]
Using this unitary, we define a rotated Bell basis as
\[
|\Phi_k\rangle = (P_k U \otimes I)\, |\Phi^+\rangle,
\]
where
\[
|\Phi^+\rangle = \frac{1}{\sqrt{2}} (|00\rangle + |11\rangle),
\]
and \( \{P_k\} = \{I, Z, X, ZX\} \) denote the Pauli operators.

This construction generates a complete orthonormal basis for the joint measurement. The measurement operators are given by
\[
M_k = |\Phi_k\rangle \langle \Phi_k| \otimes I.
\]

Following the measurement outcome \(k\), the receiver applies a corresponding correction operation defined as
\[
U_k = U P_k U^\dagger.
\]

This generalizes the standard Pauli correction scheme by allowing the correction operators to be adapted through the unitary \(U(\phi,\theta,\lambda)\). The parameters \( (\phi,\theta,\lambda) \) are optimized to maximize teleportation fidelity under noise.
\section{Adaptive Quantum Channel and Post-Processing \label{AppendixB}}

In addition to adapting the measurement basis, we introduce a generalized quantum channel acting on the output state after correction. This Post-Processing step is described by a completely positive trace-preserving (CPTP) map.

The channel is represented using a set of Kraus operators \( \{J_i\} \), satisfying
\[
\sum_i J_i^\dagger J_i = I.
\]

After the teleportation process and correction operations,the pre mentioned Post-processing channel act on the receiver’s state. The corrected state \( \rho_{\text{corr}} \) is transformed as
\[
\rho_{\text{corr}} \rightarrow \sum_j J_j \rho_{\text{corr}} J_j^\dagger,
\]

This post-processing step allows the protocol to compensate for residual noise introduced during transmission and measurement. Together, the Post-Processing Channel, Adaptive Measurement Basis and Entanglement Channel along with Adaptive Correction Operators enable a fully adaptive teleportation scheme.

The combined effect of these operations is to transform the overall teleportation protocol into a generalized quantum channel, whose structure is optimized to maximize fidelity under different noise models. The fully adaptive machine learning-tuned parameters, expressed as cubic polynomial functions for three different noise models, are presented in Tables \ref{tab:bitflip_fulladaptive}, \ref{tab:ad_fulladaptive} and \ref{tab:depol_fulladaptive}.

\onecolumngrid
\begin{table*}[t]
\caption{Cubic polynomial coefficients for Fully Adaptive Protocol - Bit-flip noise}
\label{tab:bitflip_fulladaptive}
\centering

\small
\renewcommand{\arraystretch}{1.3}
\setlength{\tabcolsep}{7pt}

\resizebox{0.86\textwidth}{0.52\textheight}{
\begin{tabular}{lcccc}
\toprule\toprule
\textbf{Parameter} & $p^3$ & $p^2$ & $p$ & Constant \\
\midrule

$\phi$ & -3.19566 & 4.47589 & -2.05096 & 0.34890 \\
$\theta$ & -2.75823 & 2.85939 & -0.80722 & 0.05076 \\
$\lambda$ & 3.07688 & -4.80518 & 2.40834 & -0.40830 \\

\midrule
$a_{\mathrm{Re}}$ & -1.25059 & 1.90817 & -0.80286 & 0.67329 \\
$a_{\mathrm{Im}}$ & 2.46227 & -3.79581 & 1.57566 & -0.18197 \\

$b_{\mathrm{Re}}$ & 0.78856 & -0.35731 & -0.13478 & 0.03160 \\
$b_{\mathrm{Im}}$ & 1.04767 & -1.49070 & 0.62490 & -0.05020 \\

$c_{\mathrm{Re}}$ & -1.42129 & 1.37480 & -0.33560 & 0.01502 \\
$c_{\mathrm{Im}}$ & -0.24185 & 0.14827 & 0.04126 & 0.00547 \\

$d_{\mathrm{Re}}$ & -1.03166 & 1.32752 & -0.38844 & 0.59382 \\
$d_{\mathrm{Im}}$ & 4.33951 & -6.40711 & 2.60280 & -0.27179 \\

\midrule
$J0_{00}^{\mathrm{Re}}$ & -1.36861 & 4.89742 & -3.70071 & 1.17404 \\
$J0_{00}^{\mathrm{Im}}$ & 1.02923 & -1.46764 & 0.38218 & 0.04519 \\

$J0_{01}^{\mathrm{Re}}$ & -1.74901 & 2.01367 & -0.58463 & 0.02676 \\
$J0_{01}^{\mathrm{Im}}$ & -0.65824 & 0.79852 & -0.21673 & -0.00525 \\

$J0_{10}^{\mathrm{Re}}$ & -1.70102 & 2.79437 & -0.95799 & 0.03271 \\
$J0_{10}^{\mathrm{Im}}$ & -1.57402 & 2.36598 & -0.87298 & 0.02934 \\

$J0_{11}^{\mathrm{Re}}$ & -1.39055 & 2.42984 & -1.22099 & 1.00784 \\
$J0_{11}^{\mathrm{Im}}$ & 0.34971 & 0.09438 & -0.47051 & 0.13094 \\

\midrule
$J1_{00}^{\mathrm{Re}}$ & 0.57527 & -0.86101 & 0.62407 & -0.11917 \\
$J1_{00}^{\mathrm{Im}}$ & -1.24827 & 0.86290 & 0.02263 & -0.01000 \\

$J1_{01}^{\mathrm{Re}}$ & -0.08687 & 0.34619 & -0.29553 & 0.01892 \\
$J1_{01}^{\mathrm{Im}}$ & -0.14088 & 0.47265 & -0.29435 & 0.02397 \\

$J1_{10}^{\mathrm{Re}}$ & 0.85304 & -0.17969 & -0.58972 & 0.04944 \\
$J1_{10}^{\mathrm{Im}}$ & -1.82525 & 2.91298 & -1.24804 & 0.08751 \\

$J1_{11}^{\mathrm{Re}}$ & -0.27887 & 0.19052 & 0.39792 & -0.10734 \\
$J1_{11}^{\mathrm{Im}}$ & 0.39307 & -1.83602 & 1.21293 & -0.08596 \\

\bottomrule
\end{tabular}
}
\end{table*}
\clearpage
\begin{table*}[t]
\caption{Cubic polynomial coefficients for Fully Adaptive Protocol - Amplitude Damping noise}
\label{tab:ad_fulladaptive}
\centering

\small
\renewcommand{\arraystretch}{1.3}
\setlength{\tabcolsep}{7pt}

\resizebox{0.86\textwidth}{0.52\textheight}{
\begin{tabular}{lcccc}
\toprule\toprule
\textbf{Parameter} & $p^3$ & $p^2$ & $p$ & Constant \\
\midrule

$\phi$ & -3.23134 & 4.29830 & -1.25000 & -0.04303 \\
$\theta$ & -0.59474 & 0.85622 & -0.25179 & -0.02840 \\
$\lambda$ & -1.45810 & 1.46173 & -0.04692 & -0.02635 \\

\midrule
$a_{\mathrm{Re}}$ & -1.44015 & 1.55017 & -0.05865 & 0.68185 \\
$a_{\mathrm{Im}}$ & 7.79834 & -11.18077 & 4.16295 & -0.25890 \\

$b_{\mathrm{Re}}$ & -1.10491 & 1.38771 & -0.37876 & -0.01559 \\
$b_{\mathrm{Im}}$ & -1.16209 & 1.32030 & -0.32948 & 0.00612 \\

$c_{\mathrm{Re}}$ & -0.17175 & 0.04318 & 0.05307 & 0.00335 \\
$c_{\mathrm{Im}}$ & -0.21119 & 0.18075 & 0.03270 & -0.03216 \\

$d_{\mathrm{Re}}$ & -0.08703 & 0.50701 & -0.79097 & 0.60592 \\
$d_{\mathrm{Im}}$ & 2.51441 & -4.26052 & 2.26491 & -0.30296 \\

\midrule
$J0_{00}^{\mathrm{Re}}$ & -0.33374 & 0.72229 & -0.44594 & 0.96429 \\
$J0_{00}^{\mathrm{Im}}$ & 1.04365 & -1.34519 & 0.38789 & 0.01670 \\

$J0_{01}^{\mathrm{Re}}$ & -0.23821 & 0.35202 & -0.10401 & -0.01349 \\
$J0_{01}^{\mathrm{Im}}$ & -0.10246 & 0.20890 & -0.12498 & 0.01963 \\

$J0_{10}^{\mathrm{Re}}$ & 0.27255 & -0.39856 & 0.11651 & 0.01341 \\
$J0_{10}^{\mathrm{Im}}$ & -0.23766 & 0.47804 & -0.29271 & 0.05147 \\

$J0_{11}^{\mathrm{Re}}$ & 0.54378 & -0.58491 & 0.05423 & 0.92038 \\
$J0_{11}^{\mathrm{Im}}$ & 3.57623 & -4.48358 & 0.93630 & 0.11963 \\

\midrule
$J1_{00}^{\mathrm{Re}}$ & -0.96251 & 2.16678 & -1.28424 & 0.09387 \\
$J1_{00}^{\mathrm{Im}}$ & -1.80414 & 2.89257 & -1.00898 & -0.07814 \\

$J1_{01}^{\mathrm{Re}}$ & -0.02540 & 0.01731 & 0.01736 & -0.00990 \\
$J1_{01}^{\mathrm{Im}}$ & 0.08683 & -0.13175 & 0.05424 & -0.00290 \\

$J1_{10}^{\mathrm{Re}}$ & -0.06653 & 0.19122 & -0.13316 & 0.02229 \\
$J1_{10}^{\mathrm{Im}}$ & -0.28118 & 0.51619 & -0.25638 & 0.02811 \\

$J1_{11}^{\mathrm{Re}}$ & 0.34447 & 0.51378 & -0.84387 & 0.10727 \\
$J1_{11}^{\mathrm{Im}}$ & -2.29937 & 3.60398 & -1.25272 & -0.05816 \\

\bottomrule
\end{tabular}
}
\end{table*}
\clearpage
\begin{table*}[t]
\caption{Cubic polynomial coefficients for Fully Adaptive Protocol- Depolarizing noise}
\label{tab:depol_fulladaptive}
\centering

\small
\renewcommand{\arraystretch}{1.3}
\setlength{\tabcolsep}{7pt}

\resizebox{0.86\textwidth}{0.52\textheight}{
\begin{tabular}{lcccc}
\toprule\toprule
\textbf{Parameter} & $p^3$ & $p^2$ & $p$ & Constant \\
\midrule

$\phi$ & 2.07663 & -1.56677 & -0.26709 & 0.19440 \\
$\theta$ & -1.65785 & 3.83734 & -2.04913 & 0.15462 \\
$\lambda$ & 0.53181 & -0.60673 & 0.28908 & -0.07386 \\

\midrule
$a_{\mathrm{Re}}$ & 2.14708 & -3.16446 & 0.66607 & 0.65088 \\
$a_{\mathrm{Im}}$ & -5.84633 & 7.55918 & -2.07774 & -0.01125 \\

$b_{\mathrm{Re}}$ & -0.16095 & 0.13377 & -0.06449 & 0.00528 \\
$b_{\mathrm{Im}}$ & -4.10010 & 5.14140 & -1.49738 & 0.07053 \\

$c_{\mathrm{Re}}$ & -0.09390 & -0.51717 & 0.49244 & -0.05162 \\
$c_{\mathrm{Im}}$ & 1.35993 & -1.97570 & 0.53961 & -0.01903 \\

$d_{\mathrm{Re}}$ & 2.13406 & -2.99971 & 0.80368 & 0.63719 \\
$d_{\mathrm{Im}}$ & 2.45063 & -2.75896 & 0.67344 & -0.01700 \\

\midrule
$J0_{00}^{\mathrm{Re}}$ & 3.21564 & -3.23014 & -0.85101 & 1.00296 \\
$J0_{00}^{\mathrm{Im}}$ & -0.42514 & 0.91070 & -0.57524 & 0.10961 \\

$J0_{01}^{\mathrm{Re}}$ & -0.20289 & 0.36513 & -0.17663 & 0.01555 \\
$J0_{01}^{\mathrm{Im}}$ & 0.00641 & 0.02784 & -0.04757 & 0.01604 \\

$J0_{10}^{\mathrm{Re}}$ & -0.02901 & -0.62196 & 0.40364 & -0.02699 \\
$J0_{10}^{\mathrm{Im}}$ & -2.26739 & 3.37979 & -1.15590 & 0.05136 \\

$J0_{11}^{\mathrm{Re}}$ & -0.54610 & 0.67453 & -0.34442 & 0.94700 \\
$J0_{11}^{\mathrm{Im}}$ & 2.48892 & -3.29262 & 1.25133 & -0.09061 \\

\midrule
$J1_{00}^{\mathrm{Re}}$ & -0.64919 & 0.86281 & -0.19275 & -0.05214 \\
$J1_{00}^{\mathrm{Im}}$ & -1.46790 & 2.57694 & -1.29761 & 0.16067 \\

$J1_{01}^{\mathrm{Re}}$ & -0.02786 & 0.05647 & -0.02175 & -0.00048 \\
$J1_{01}^{\mathrm{Im}}$ & -0.22180 & 0.32368 & -0.14133 & 0.01963 \\

$J1_{10}^{\mathrm{Re}}$ & -4.38474 & 6.95693 & -2.51534 & 0.08025 \\
$J1_{10}^{\mathrm{Im}}$ & 1.56049 & -0.83828 & -0.47893 & 0.03398 \\

$J1_{11}^{\mathrm{Re}}$ & 0.99533 & -1.44058 & 0.46795 & -0.03724 \\
$J1_{11}^{\mathrm{Im}}$ & -1.02098 & 3.11069 & -2.05930 & 0.23449 \\

\bottomrule
\end{tabular}
}
\end{table*}
\clearpage
\twocolumngrid
\bibliography{apssamp}
\end{document}